\documentclass[%
 reprint,%
 amssymb, amsmath,%
 aip,jap,%
]{revtex4-1}
\usepackage{graphicx}
\usepackage{dcolumn}
\usepackage{bm}
\usepackage{amsmath}
\usepackage{tabularx}
\usepackage[colorlinks]{hyperref}
\hypersetup{urlcolor=blue, citecolor=black, colorlinks=true, linkcolor=blue} %Colours hyperlinks in blue, but this can be distracting if there are many links.
\usepackage{url}
\begin{document}
\preprint{AIP/123-QED}
\title{Anti-site disorder and improved functionality of Mn$_{2}$Ni{\it X}
({\it X}= Al, Ga, In, Sn) inverse Heusler alloys}
\author{Souvik Paul} %
\affiliation{Department of Physics, Indian Institute of Technology
Guwahati, Guwahati, Assam 781039, India} %
\author{Ashis Kundu} %
\affiliation{Department of Physics, Indian Institute of Technology
Guwahati, Guwahati, Assam 781039, India} %
\author{Biplab Sanyal}
\affiliation{Department of Physics and Astronomy,
Uppsala University, Box 516, 75120 Uppsala, Sweden}
\author{ Subhradip Ghosh}
\email{subhra@iitg.ernet.in}
\affiliation{Department of Physics, Indian Institute of Technology
Guwahati, Guwahati, Assam 781039, India} %

\date{\today}

\begin{abstract}

Recent first-principles calculations have predicted Mn$_{2}$Ni{\it X} ({\it X}=Al, Ga, In, Sn)
alloys to be magnetic shape memory alloys. Moreover, experiments on Mn$_{2}$NiGa
and Mn$_{2}$NiSn suggest that the alloys deviate from the perfect inverse Heusler
arrangement and that there is chemical disorder at the sublattices with tetrahedral
symmetry. In this work, we investigate the effects of such chemical disorder on
phase stabilities and magnetic properties using first-principles electronic
structure methods. We find that except Mn$_{2}$NiAl, all other alloys show signatures
of martensitic transformations in presence of anti-site disorder at the
sublattices with tetrahedral symmetry. This improves the possibilities of realizing
martensitic transformations at relatively low fields and the possibilities of obtaining
significantly large inverse magneto-caloric effects, in comparison to perfect inverse Heusler arrangement of atoms.
We analyze the origin of such improvements
in functional properties by investigating electronic structures and magnetic
exchange interactions.

\end{abstract}

\maketitle
\section{Introduction}

For last couple of decades, magnetic shape memory alloys (MSMAs) have been
fascinating the researchers because these materials combine different
degrees of freedom like structural, magnetic, elastic, caloric etc.,
in a single phase. Coupling of micro-structure with magneto-crystalline
anisotropy enables increase of
magnetic field-induced strain (MFIS), which makes this class
of smart materials suitable for micro-mechanical automotive sensors and actuators
applications. Other interesting multifunctional properties exhibited by this
class of materials are conventional and inverse magneto-caloric
effect \cite{magcal1,magcal2,magcal3,toprev},
barocaloric \cite{barocal}, elastocaloric \cite{elascal},
giant magnetoregistance \cite{magres1,magres2}, exchange bias \cite{exbias},
kinetic arrest \cite{ka}, spin-glass \cite {spglass} and strain-glass \cite{stglass}.
In the heart of these applications is the magnetic-field driven martensitic transformation.
The magnitude and sign of the magnetization difference ($\Delta M$) between the high
temperature austenite phase and the low temperature martensite phase is a key quantity
responsible for the functional properties mentioned above. For example,
Zeeman energy plays a crucial role in realizing the martensitic transformation
in ferromagnetic shape memory alloys. An externally applied magnetic field
performs two different tasks inside MSMA. Firstly, if the direction of the
applied field is different from the easy magnetic axis of the material, the
field tries to rotate the magnetization direction along it against the
force associated with magnetic anisotropy. Secondly, the field generates
driving force across the twin boundaries between martensitic variants,
the associated energy is known as Zeeman energy. Since the energy required to move martensitic
domains is lower than the magnetocrystralline anisotropy energy in MSMA,
the pressure created by Zeeman energy increases the volume fraction of
favorably oriented martensitic variants
and lowers the
energy of the product phase. It has been found that
a large $\Delta M$ between the two phases in the
presence of moderate external magnetic field $H$ , i.e., the Zeeman term $\Delta M.H$,
drives the motions of the martensitic domains facilitating the martensitic
transformation \cite{co-mnniga,fe-mnniga}. On the other hand, the sign of $\Delta M$ is
related to the conventional or inverse magnetocaloric effect \cite{toprev}. In case
of the former, a magnetic field will cause a decrease in entropy when applied isothermally
and an increase in temperature when applied adiabatically. In case of the later, an increase
in entropy will effect a decrease of temperature leading to ``magnetic cooling'', a phenomenon
which can be exploited for green technology in refrigeration. For the inverse magnetocaloric
effect, the magnetization in the martensitic phase has to be lower than that in the austenite
phase. Recent research on MSMAs thus, have focused on exploring materials where $\Delta M$ can
be suitably tuned in order to obtain large Zeeman energy with a small field and/or a substantial
inverse magneto-caloric effect.

A decade ago, Ni$_{2}$MnGa was discovered as the first promising MSMA \cite{nimnga1,nimnga2}.
However, Ni$_{2}$MnGa in the stoichiometric composition of 2:1:1 shows only conventional
magneto-caloric effect. Moreover, $\Delta M$ for the system is rather small. Other
materials in the Ni$_{2}$Mn{\it X} ({\it X}= Al, In, Sn)
series do not even posses shape memory properties
at their stoichiometric compositions. A lot of research has taken place since then to obtain
shape memory and/or inverse magneto-caloric effect in Ni$_{2}$MnX systems by tuning the electron
to atom ($e/a$) ratio by either changing concentration  ratio of Ni,Mn and X \cite{toprev}
or by doping with another element \cite{entel1,entel2}. These approaches led to the investigations
into the potentials of Mn$_{2}$Ni{\it X} ({\it X}= Al, Ga, In, Sn) systems.
%Recently, a spin valve like magnetoresistance behavior is observed in bulk
%Mn$_{2}$NiGa SMA at room temperature\cite{spinvlv}.
The exploration in Mn$_{2}$Ni{\it X}
materials was initiated in 1987 by Helmholdt {\it et al.} with their investigation
in the structural and the magnetic properties of Mn$_{2}$NiSn form Neutron
diffraction data \cite{mn2nisnexp}. Their study concluded that unlike the
Ni$_{2}$Mn{\it X} materials which form the usual Heusler structure,
Mn$_{2}$NiSn crystallizes in inverse Heusler structure. The later structure
forces one of the tetrahedral Mn atoms to interchange its position with
octahedral Ni atom making the two Mn atoms nearest neighbors.
The next significant step was investigations into the shape memory related
properties of Mn$_{2}$NiGa, conducted by Liu {\it et al.} \cite{mn2niga}.
The material exhibits
an excellent two-way shape memory effect where the martensitic transformation temperature
(T$_{M}\sim$ 270 K) is close to room temperature and the Curie temperature
(T$_{C}\sim$ 588 K) is quite high \cite{mn2niga}.
The last two parameters of Mn$_{2}$NiGa are
better than those of Ni$_{2}$MnGa. This study hinted at
the better functional properties of this material compared to the prototype and thus,
generated immense interests into investigations of the Mn$_{2}$Ni{\it X} materials with
the possibilities of discovering new MSMAs suitable for practical applications.
It is noteworthy to mention that Mn$_{2}$NiSn also has very high Curie
temperature (T$_{C}\sim$ 530 K) \cite{lakshmi}.
X-ray diffraction experiments confirmed that Mn$_{2}$NiGa also forms an
inverse Heusler structure in the austenite phase. Experimental reports
on the structure and shape memory property of
Mn$_{2}$NiAl and Mn$_{2}$NiIn are not yet available.
However theoretical calculations on Mn$_{2}$Ni{\it X} confirm that all
the materials exhibit shape memory effect and
prefer inverse Heusler structure \cite{paul}.
The direct consequence for crystallizing in this structure is lower saturation
magnetization due to opposite and unequal moments of nearest neighbor Mn atoms,
as compared to prototype Ni$_{2}$MnGa, where all the constituent atoms couple
ferromagnetically. For Mn$_{2}$NiAl and Mn$_{2}$NiGa, total magnetization reported
is $\sim$1 $\mu_{B}$ and for Mn$_{2}$NiIn and Mn$_{2}$NiSn those are $\sim$0.5 $\mu_{B}$ \cite{paul},
while Ni$_{2}$MnGa posses a high moments of $\sim$4 $\mu_{B}$. Since the MFIS is
directly proportional to the total magnetization, lowering of the later would results
in a smaller MFIS in Mn$_{2}$NiGa ($\sim$4$\%$), whereas largest MFIS are observed in
nearly stoichiometric composition of NiMnGa MSMA is about 10$\%$ \cite{nimnga1,nimnga2}.
On the contrary,
Helmholdt {\it et al.} observed a higher value of total magnetization,
nearly 2.5 $\mu_{B}$, in cubic austenite phase of Mn$_{2}$NiSn \cite{mn2nisnexp}.
They showed that within the inverse Heusler structure, there is significant anti-site
disorder among the Ni and Mn atoms at the tetrahedral positions. Recent Neutron diffraction
studies \cite{brown,spinvlv} and first principles Density Functional Theory (DFT) calculations
on Mn$_{2}$NiGa \cite{aparch} and Mn$_{2}$NiSn \cite{pauljpcm} also confirmed the presence
of anti-site disorder in these materials. DFT calculations conclusively showed that such
anti-site disorder is responsible for a significantly large magnetic moment in the
high temperature phase. The calculations also showed that Mn$_{2}$NiGa would exhibit
inverse magneto-caloric effect \cite{aparch}.

Motivated by these discoveries, in this communication, we perform DFT calculations to
investigate the magnetic properties
of the four Mn$_{2}$Ni{\it X} alloys which are potential MSMAs, in crystal structures relevant to
the austenite and martensite phases, with and without anti-site disorder. We specifically
looked at the trends in $\Delta M$ across the series. We analyze the results
by computing the electronic structures and the inter-atomic exchange interactions for each
of the cases. The outcome of this systematic study indicates that the Mn$_{2}$Ni{\it X} alloys
may have better functionalities than the prototype MSMAs, which is
driven by the anti-site disorder inherent to these systems, an important feat relevant in todays
technology. The paper is organized as follows: In Section II, we discuss the computational details.
Section III contains the results and subsequent discussions on the magnetic properties, the electronic
structure and the inter-atomic exchange interactions. Summary and importance of this work are
described in the conclusions section.

\section{Computational Details}

The electronic structures of the materials concerned have been calculated
using Full-potential based Spin-Polarized Relativistic Korringa-Kohn-Rostoker
(SPR-KKR) Green's function method \cite{kkr,sprkkr}.
The Local Spin Density Approximation
(LSDA) as parameterized by Vosko-Wilk-Nusair (VWN) was used as the
exchange-correlation part of the potential to solve the Kohn-Sham equation \cite{vwn}.
The angular momentum cut-off to the plane wave was taken to be $\ell_{max}$=3.
The Brillouin zone integrations have been carried out on a uniform
24$\times$24$\times$24 {\it k}-mesh. The Green's function was calculated for
30 complex energy points distributed exponentially on a semicircular contour.
The energy convergence criterion was set to 10$^{-6}$ Ry for the self-consistent
cycles. The Coherent Potential Approximation (CPA)
was used to incorporate the effects of disorder \cite{cpa}.

The magnetic pair exchange interactions have been calculated
with multiple scattering Green function formalism as implemented in SPR-KKR code \cite{sprkkr}.
Spin fluctuation theories for metals map the complicated itinerant electron systems
onto an effective Heisenberg Hamiltonian having the classical spins as
\begin{eqnarray}
H_{eff}= -\sum_{\mu,\nu}\sum_{i,j}
J^{\mu\nu}_{ij}
\mathbf{e}^{\mu}_{i}
.\mathbf{e}^{\nu}_{j}
\end{eqnarray}
The indices $\mu$ and $\nu$ represent different sublattices,
\emph{i} and \emph{j} denote the
atomic positions and $\mathbf{e}^{\mu}_{i}$ is the unit vector
along the direction of magnetic moments at site \emph{i} belonging to sublattice $\mu$.
The size of the magnetic moments are included in the exchange
interaction parameter $J^{\mu \nu}_{ij}$.
The exchange parameters are computed from energy difference due to
the small orientation of a pair of spins resulting a perturbation
in spin-density which within the formulation of Lichtenstein {\it et al.}
\cite{jij} based on magnetic force theorem \cite{mft} takes the following form
\begin{eqnarray}
J^{\mu\nu}_{ij}= \frac{1}{4\pi}\int^{E_{F}}_{-\infty}
d\epsilon \Im \mathrm{Tr}
(\Delta_{i}\hat{T}^{ij}_{\sigma}
\Delta_{j}\hat{T}^{ji}_{\sigma\prime})
\end{eqnarray}
where $\Delta_i$=
$\hat{t}^{-1}_{i\sigma}$-$\hat{t}^{-1}_{i\sigma\prime}$,
$\sigma$ is the spin index, $\hat{t}$ is the single site scattering
matrix and $\hat{T}$ is the scattering path operator related to the
off-diagonal elements of the Green's function. $\mathrm{Tr}$ is the
trace over the orbital indices of the scattering matrix. Positive
(negative) values for J$^{\mu\nu}_{ij}$ indicate ferromagnetic
(antiferromagnetic) coupling between atoms $i$ and $j$.

\begin{table}[h]
    \caption{Sublattice occupancies corresponding to the configurations used. The details
            are described in the text.}
    \begin{center}
        \begin{tabular}{lcccc}
        \hspace{0.37in} & \hspace{0.67in} & \hspace{0.67in} & \hspace{0.37in} &\hspace{0.47in}$\,$\\
        Configuration & & Sublattice& &  \\
        \hline
        & $(000)$ & $(\frac{1}{2} \frac{1}{2} \frac{1}{2})$ & $(\frac{1}{4} \frac{1}{4} \frac{1}{4})$ & ($\frac{3}{4}$$\frac{3}{4}$$\frac{3}{4}$) \\
        \hline
        {\bf OC,OT} & MnI & Ni & MnII & X  \\
        \hline
        {\bf DC,DT} & (MnI$_{0.5}$)Ni$_{0.5}$ & (MnI$_{0.5})$Ni$_{0.5}$ &MnII & X \\
        \hline
        \end{tabular}
    \end{center}
\end{table}

\begin{figure}[h]
\includegraphics[width=8.5cm]{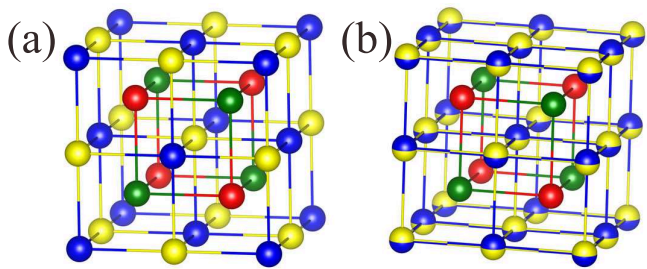}
\caption{(Color online) The crystal structure of Mn$_{2}$Ni{\it X} in (a) {\bf OC} and in (b) {\bf DC} configurations.
The blue, green, yellow and red spheres represent MnI, MnII, Ni and {\it X} atoms, respectively. The two-toned
spheres (blue and yellow) represent sublattices with Ni-MnI alloys.}
\end{figure}

Table I shows different configurations used in this work. Space group $Fm\bar{3}m$ has
been used for the cubic austenite phase and $Fmmm$ for the tetragonal
martensitic phase. The configurations without anti-site disorder, i.e., ones corresponding
to the perfect inverse Heusler arrangement have been referred to as {\bf OC} (Ordered Cubic) and {\bf OT}
(Ordered Tetragonal), while the ones with anti-site disorder have been referred to as {\bf DC}
(Disordered Cubic) and {\bf DT} (Disordered Tetragonal). In cases of {\bf OC} and {\bf OT}, the sublattices with
octahedral symmetries are occupied by the {\it X} element and one of the Mn atoms (referred to as MnII), while the
sublattices with tetrahedral symmetries are occupied by Ni and the other Mn atom (referred to as MnI). In cases
of {\bf DC} and {\bf DT}, the tetrahedral sublattices have anti-site disorder and hence they consist of binary
alloys of MnI and Ni, MnI$_{x}$Ni$_{1-x}$. The concentration $x$ is found to vary depending upon the system and the
experiment \cite{brown,mn2nisnexp}; hence we have considered $x$ to be 0.5 uniformly across all systems.
The reasons for choosing $x=0.5$ are as follows: first, the neutron diffraction results on Mn$_{2}$NiGa suggested
that the occupancies of sublattices in the inverse Heusler structure in the system would be like the
configuration considered here \cite{brown}; for Mn$_{2}$NiSn, although the value of $x$ suggested by the experiment
was slightly different \cite{mn2nisnexp}, the results of first-principles calculations \cite{pauljpcm} showed that
there are small differences between the magnetizations (over a sufficiently large range of volume) calculated
with the system being in the configuration of Table I and that calculated using the configuration suggested by the
experiment, and second, the choice of $x=0.5$ maximizes the anti-site disorder on the tetrahedral sites. For example,
$x=0.4$ at the MnI site and $x=0.6$ at the Ni site produces the same configuration as the one with $x=0.6$ at the MnI site and $x=0.4$
at the Ni site since the MnI and Ni sites are crystallographically equivalent. Moreover, $x>0.5$ at any of the two sites produces
a configuration closer to the ``ordered'' ({\bf OC} or {\bf OT}) one.
Figure I shows the schematic representation of the {\bf OC} and {\bf DC} structures.

However, before proceeding further with the {\bf DC} and {\bf DT} configurations, one must make sure that
\[F_{DC} < F_{OC}, \enskip \enskip \enskip \enskip F_{DT} < F_{OT} \]
Here F$_{DC}$, F$_{OC}$, F$_{DT}$ and $F_{OT}$ refer to the free energies in the four configurations respectively. These
equations make sure that the ``anti-site disordered'' configurations are the thermodynamically favorable ones over the
``ordered'' ones, thus, making the justifications for analyzing results obtained with the ``anti-site disordered'' configurations
any further. On top of this, one also has to make sure that the energy of the tetragonal phase in the ``anti-site disordered''
configuration is lower than that of the cubic phase in the same configuration, for each of these systems. Unless this occurs,
further investigations into the systems exploring improved functionalities, based upon magnetizations in the austenite and
in the martensite phases, would be useless.

In what follows, we compute the following quantities
\begin{eqnarray*}
\Delta E_{str}^{O/D} &=& E_{tet}^{O/D}-E_{cub}^{O/D} \\
\Delta F_{ord}^{C} &=& F_{OC}-F_{DC} \\
\Delta F_{ord}^{T} &=& F_{OT} -F_{DT}
\end{eqnarray*}
$\Delta E_{str}^{O}$ and $\Delta E_{str}^{D}$ refer to the energies required for structural transition from the cubic to the tetragonal phase when the
system is in ``ordered'' configuration {\bf O} and in ``anti-site disordered'' configuration {\bf D}, respectively. A negative value of
$E_{str}^{O/D}$ means that the tetragonal phase is energetically lower than the cubic phase in the {\bf O/D} configuration and that the
martensitic transformation is possible. $\Delta F_{ord}^{C}$ and $\Delta F_{ord}^{T}$ are the free energies of the configuration {\bf O}
with reference to those of the configuration {\bf D} in the cubic ($C$) and in the tetragonal ($T$) phases, respectively. The free energy expression considered here is,
\begin{eqnarray*}
F=E+\frac{k_{B}T}{N} \sum_{i} x_{i} ln x_{i} +\left(1-x_{i} \right) ln \left(1-x_{i} \right)
\end{eqnarray*}
$E$ is the electronic energy per atom, $k_{B}$ is the Boltzmann constant, $T$ is the temperature, $N$ is the number of atoms, $i$ is the sublattice
index and $x_{i}$ is the concentration of sublattice $i$. Here we have considered only the contribution of the configurational part to
the entropy and neglected the effects of the lattice vibrations and electronic temperatures. At ambient conditions, the effect of the
electronic temperature are negligible. The
contribution from lattice vibration to the free-energy difference for alloys with different site-occupation configurations
can be estimated approximately from the high-
temperature expansion of the phonon free energy $\Delta F_{ph} \sim 3kT \left(\Delta \Theta/ \Theta \right)$ \cite{debye}.
In the simplest approximation, the
Debye temperatures $\Theta $ are proportional to $\sqrt rB$ \cite{debye1}, where $r$ is
the Wigner-Seitz radius and $B$ is the bulk modulus. For the systems considered here, the lattice
constants and hence the Wigner-Seitz radii differ only slightly between the {\bf O} and the {\bf D} configurations; same
happens for Bulk moduli. Consequently, the contributions from vibrational part to the entropy are orders of magnitude
smaller than the electronic contributions. In this paper, we have calculated the contributions from configurational
entropy to the free energies only at $T=300$ K.

In materials with chemical disorder, particularly in systems where constituents have large size differences such as Mn$_{2}$NiIn
and Mn$_{2}$NiSn, there can be substantial relaxations of the local bonds. For Mn$_{2}$NiSn, the impact of such relaxations were
found to be non-negligible \cite{pauljpcm}. Since incorporation of the relaxations is not possible within the framework of the
KKR-CPA method, we compute $\Delta E_{str}^{O/D}, \Delta F_{ord}^{C}, \Delta F_{ord}^{T}$ by VASP plane wave code employing the
Projector Augmented Wave approach \cite{vasp1,vasp2}. In order to mimic the chemical disorder for ``anti-site disordered'' configurations,
we consider 64 atom supercells with MnI and Ni atoms occupying the sites with tetrahedral symmetry randomly. We also construct a 64
atom ``Special Quasi-random Structure'' (SQS) \cite{sqs} which supposedly simulates the environments around an atom in a chemically
disordered alloy better as the SQS is generated by matching the maximum number of correlation functions to their exact values in the
real disordered alloy. The SQS structures used in this work are generated with the `Alloy Theory Automated Toolkit' (ATAT) package
\cite{atat}. 50 pairs and 30 triplets having correlation functions exactly equal to those for a real disordered alloy with the
same composition were chosen to construct the SQS. The total energies were calculated with both the supercell and the SQS structure.
In each case, all atoms were relaxed keeping the volume fixed. A large basis was used with a plane wave cut-off of 450 eV. The exchange-correlation functional used was the same as that used in KKR calculations. Convergences of the electronic structures were assumed when changes between two consecutive steps were less than $10^{-5}$ eV. Atomic relaxations were carried out until all the forces were less
than $10^{-3}$ eV/\AA. A $k$-point mesh consisting of at least 9 $k$ points in the irreducible part of the Brillouin Zone was
considered and was sufficient for convergences of total energies and forces.

\section{Results and Discussions}

\subsection{Structural parameters of Mn$_{2}$NiX}

The experimental information on structural parameters for cubic
phase are available for Mn$_{2}$NiGa \cite{brown,mn2niga} and Mn$_{2}$NiSn \cite{mn2nisnexp}
only, while the information on the structural parameters in the tetragonal phase is
available for Mn$_{2}$NiGa alone. In this work, we have, therefore, used the
experimental lattice parameters of Mn$_{2}$NiGa in both phases and of Mn$_{2}$NiSn
in the cubic phase. The lattice parameters for the other two alloys in both phases and
for Mn$_{2}$NiSn in tetragonal phase were calculated by first-principles DFT. The results are tabulated in Table II.

\begin{table}[!htbp]
    \caption{The lattice parameters for Mn$_{2}$NiX systems used in this work.a$_{cubic}$ and a$_{tet}$ are the
lattice constants in the cubic and tetragonal phases respectively, (c/a)$_{tet}$ is the global minima in the tetragonal phase.}
    \begin{center}
        \begin{tabular}{|c|c|c|c|} \hline
        Systems & lattice constant  & lattice constant &(c/a)$_{tet}$ \\
                & a$_{cubic} (\AA)$ & a$_{tet} (\AA)$ & \\ \hline
        Mn$_{2}$NiAl &5.57 &3.71 &1.20\\ \hline
        Mn$_{2}$NiGa &5.90 &3.92 &1.21\\ \hline
        Mn$_{2}$NiIn &5.96 &3.91 &1.26\\ \hline
        Mn$_{2}$NiSn &6.10 &4.06 &1.20\\ \hline
        \end{tabular}
    \end{center}
\end{table}

The results obtained for {\bf OC} and {\bf OT} configurations using SPRKKR and VASP codes
were nearly identical, with a maximum difference less than $1 \%$. 
The same set of lattice parameters have been used for ``ordered'' and ``disordered'' configurations
since it was found out that they hardly change from one configuration to another.
It is worth mentioning that the lattice constants of Mn$_{2}$NiGa
and Mn$_{2}$NiSn, in their cubic phases, calculated by DFT are 5.64 \AA and 5.92 \AA respectively. The
DFT results differ by $3.5-4 \%$ compared to the experimentally obtained lattice parameters, presumably due
to the LSDA functionals. We found that this discrepancy is consistent irrespective of the basis set used.

\subsection{Stabilities of different configurations}

In Table III, we show our results on $\Delta E_{str}^{O/D}, \Delta F_{ord}^{C}$ and $\Delta F_{ord}^{T}$,
obtained after relaxations of atomic positions in the supercell and in the SQS structure calculated
by the VASP code. We note that apart from qualitative agreement on the trends, the quantitative agreement
is also close between the two different supercells simulating the ``anti-site disordered'' configurations. We find that $\Delta E_{str}^{D} > 0$
for Mn$_{2}$NiAl implying that in the ``anti-site disordered '' configuration, the martensitic transformation does not take place.
Therefore, this system in the disordered configuration would not be suitable for the functionalities related to martensitic
phase transformations and thus we exclude this system from the rest of our discussions. For the other three systems,
$\Delta E_{str}^{D} < 0$ along with $\Delta F_{ord}^{C/T} < 0$ implying that the ``anti-site disordered'' configuration ({\bf D}) is
thermodynamically favorable over the ordered configuration {\bf C} for each of these systems irrespective of the crystal
structure,and that the martensitic transformation takes place in all of them with the configuration {\bf D}.

These results, thus justify the choice of the particular disordered configuration considered in this work and further investigations
into its impacts on the functional properties can be carried out.

\begin{table*}[!htbp]
    \caption{Compilation of $\Delta E_{str}^{O/D}, \Delta F_{ord}^{C}, \Delta F_{ord}^{D}$
    for Mn$_{2}$Ni{\it X} materials calculated for 64 atom supercell and 64 atom SQS.
    Calculations
    are done by VASP-PAW code.}
    \begin{center}
        \begin{tabularx}{\linewidth}{|c|X|X|X|X|X|X|X|X|X|} \hline
        &  \multicolumn{4}{|c|}{64 atom supercell} & \multicolumn{4}{|c|}{64 atom SQS}  \\ \hline
        Systems  & $\Delta F_{ord}^{C}$ &$\Delta F_{ord}^{T}$ &$\Delta E_{str}^{O}$ &$\Delta E_{str}^{D}$ &$\Delta F_{ord}^{C}$ &$\Delta F_{ord}^{T}$&$\Delta E_{str}^{O}$ &$\Delta E_{str}^{D}$ \\
                &(meV/ &(meV/ &(meV/ &(meV/ &(meV/ &(meV/ &(meV/ &(meV/  \\
		&atom) &atom) &atom) &atom) &atom) &atom) &atom) &atom)  \\ \hline
                 %& \multicolumn{5}{|c|}{Cubic} & \multicolumn{5}{|c|}{tetragonal}\\ \hline
        Mn$_{2}$NiAl  &32&9.7 &-8.11 &14.6 &37 & 8.9&-8.2 &19.9 \\ \hline
        Mn$_{2}$NiGa  &10.0&3.5 &-22.4 &-15.8 & 9 & 3.6 &-21.6 &-16.8 \\ \hline
        Mn$_{2}$NiIn &31 &11.4 &-20.6 &-1.5 &28.6 & 12.1 &-19.9 &-3.4 \\ \hline
        Mn$_{2}$NiSn &30&27.7 &-5.83 &-3.2 & 32.4 &37.4 &-4.2 &-2.2 \\ \hline
        \end{tabularx}
    \end{center}
\end{table*}

\subsection{Magnetic moments of Mn$_{2}$Ni{\it X}}

In what follows, we compute the magnetic moments, the electronic structures, the
exchange interactions and the Curie temperatures for Mn$_{2}$NiGa, Mn$_{2}$NiIn and
Mn$_{2}$NiSn in {\bf OC}, {\bf OT}, {\bf DC} and {\bf DT} configurations using the
KKR-CPA method.
Previous works \cite{mn2niga,paul,pauljpcm} have established that the ground state
in all four systems considered here is ferrimagnetic. Accordingly,
the starting spin configurations
between two nearest neighbor Mn atoms in all four configurations considered are kept
anti-parallel.
In the {\bf DC} and {\bf DT} configurations, the MnI atoms occupying the sublattices with
tetrahedral symmetry are considered to be parallel in the beginning as they are not
the nearest neighbors. These starting configurations lead to the lowest energy states.
The calculated total and partial magnetic moments of Mn$_{2}$Ni{\it X} materials, excluding Mn$_{2}$NiAl,
in all four configurations are summarized in Table IV.

\begin{table*}[!htbp]
    \caption{Total (M$_{tot}$) and partial (M$_{i}$) moments of the constituent atoms
    for Mn$_{2}$Ni{\it X} materials in four different configurations.
    $\Delta$M represents the total moment in martensite phase with respect to that
    in the austenite phase for a given sub-lattice occupancy(either {\bf O} or {\bf D}).}
    \begin{center}
        \begin{tabular}{|c|c|c|c|c|c|c|c|c|c|c|c|c|c|} \hline
       % & \multicolumn{5}{|c|}{Cubic} & \multicolumn{5}{|c|}{Tetragonal} & \\ \hline
        Systems  & Config. &M$_{tot}$ &M$_{MnI}$ &M$_{MnII}$ &M$_{Ni}$ &M$_{X}$ &Config. &M$_{tot}$ &M$_{MnI}$ &M$_{MnII}$ &M$_{Ni}$ &M$_{X}$ &$\Delta$M\\
                & &($\mu_{B}$/f.u.) &($\mu_{B}$) &($\mu_{B}$) &($\mu_{B}$) &($\mu_{B}$)  & &($\mu_{B}$/f.u.) &($\mu_{B}$) &($\mu_{B}$) &($\mu_{B}$) &($\mu_{B}$) &($\mu_{B}$) \\ \hline
                 %& \multicolumn{5}{|c|}{Cubic} & \multicolumn{5}{|c|}{tetragonal}\\ \hline
        Mn$_{2}$NiGa & {\bf OC} &1.26 &-2.27 &3.17 &0.35 & 0.01 & {\bf OT} &1.12 &-2.27 &3.05 &0.33 & 0.01 &-0.01\\
                     & {\bf DC} &2.13 &-1.42 &3.15 &0.40 & 0.00 & {\bf DT} &1.42 &-2.02 &3.07 &0.36 & 0.01 &-0.71\\ \hline
        Mn$_{2}$NiIn & {\bf OC} &1.02 &-2.42 &3.17 &0.27 &-0.00 & {\bf OT} &1.00 &-2.30 &3.02 &0.28 & 0.00 &-0.02\\
                     & {\bf DC} &2.09 &-1.40 &3.15 &0.35 &-0.01 & {\bf DT} &1.22 &-2.10 &3.01 &0.31 & 0.00 &-0.86\\ \hline
        Mn$_{2}$NiSn & {\bf OC} &0.71 &-2.74 &3.33 &0.11 & 0.01 & {\bf OT} &0.38 &-2.84 &3.19 &0.04 &-0.01 &-0.33\\
                     & {\bf DC} &2.38 &-1.26 &3.40 &0.23 & 0.01 & {\bf DT} &1.68 &-1.96 &3.34 &0.29 & 0.02 &-0.70\\ \hline
        \end{tabular}
    \end{center}
\end{table*}

\begin{table*}[!htbp]
        \caption{The inter-atomic distances (in $\AA$) between the magnetic
        atoms for Mn$_{2}$Ni{\it X} systems in cubic and tetragonal phases.}
        \begin{center}
            \begin{tabular}{|c|c|c|c|c|c|c|} \hline
                Systems & \multicolumn{3}{|c|}{Cubic} & \multicolumn{3}{|c|}{Tetragonal}\\ \hline
                Mn$_{2}$Ni{\it X} &MnI-MnII &Ni-MnI &MnII-Ni &MnI-MnII &Ni-MnI &MnII-Ni \\ \hline
                Mn$_{2}$NiGa      &2.56      &2.95   &2.56     &2.58      &2.77   &2.58 \\
                Mn$_{2}$NiIn      &2.58      &2.98   &2.58     &2.61      &2.76   &2.61 \\
                Mn$_{2}$NiSn      &2.64      &3.05   &2.64     &2.66      &2.87   &2.66 \\ \hline
        \end{tabular}
    \end{center}
\end{table*}
The central quantity related to multi-functionalities in the present context, $\Delta M$ is
defined as $\Delta M= M_{tot}^{martensite}-M_{tot}^{austenite}$, the total moment in the martensite phase
with respect to that in the austenite phase for a given occupancy of the sublattices. Calculated values of
$\Delta M$ are also presented in Table IV.
The results show the following trends:

(i) The magnetic moments in the martensitic phase is lower than that in the austenite
phase for both ``ordered'' and ``disordered'' configurations and for all three alloys under
consideration, making $\Delta M <0$.

(ii) For the ``disordered'' configurations,that is, both {\bf DC} and {\bf DT}, the magnetic
moments are substantially higher than their ``ordered'' counterparts, i.e., {\bf OC} and {\bf OT}
respectively. The increment is significantly greater in austenite phases than that in the
martensitic phases. The total moment increases by $69 \%,104 \%$
and $235 \%$ for Mn$_{2}$NiGa, Mn$_{2}$NiIn and Mn$_{2}$NiSn respectively as configuration changes from {\bf OC} to
{\bf DC}.
In the martensitic phases, the increase in moments as configuration changes from {\bf OT} to {\bf DT}
are $27 \%, 22 \%$ and $ 342 \%$ for Mn$_{2}$NiGa, Mn$_{2}$NiIn and Mn$_{2}$NiSn
respectively. This results in a huge gain in $\Delta M$ in the ``anti-site disordered'' configurations
over that in the ``ordered'' configurations.

The behavior of the magnetic moments across crystal structures, configurations and materials, are
driven by the crystallographically inequivalent Mn atoms. The results suggest that the ferrimagnetic
ground states are achieved due to the anti-parallel alignments of the MnI and MnII atoms. In the ``ordered''
configurations, the magnitude of the moments compensate each other substantially resulting in relatively
low moments. Drastic changes are observed in cases of anti-site disorder. The substantially high moments
in {\bf DC} and {\bf DT} configurations are driven by considerable quenching of Mn moments at the sites
where they form alloys with Ni. The moments of the other Mn atom remain nearly unaltered. The Ni moments, on
the other hand, increases significantly, bringing in an increase in the net moments. The changes are more
dramatic in the austenites where the MnI moment changes between $37-54\%$ as compared to changes between $8-30 \%$
in the martensites.

The significant changes in the MnI moments driven by the anti-site disorder coupled with a substantially larger
changes in the total moments in the austenites lead to larger $\Delta M$ when anti-site disorder affects the
sublattices with tetrahedral symmetry in the inverse Heusler structure. This is a significant result as it shows that
all three Mn$_{2}$NiX alloys may exhibit inverse magneto-caloric effects and a large Zeeman energy can be
achieved with relatively small magnetic field making them attractive from the point of view of shape memory
applications. It may be noted that the prototype MSMA Ni$_{2}$MnGa has $\Delta M > 0$ and the magnitude of $\Delta M$ is
only ~ $0.2 \mu_{B}$ \cite{karin}. Thus, Mn$_{2}$NiX alloys turn out to be MSMAs with better functional parameters
even with the stoichiometric composition of 2:1:1. In the sections IIID and IIIE, we explore the origin of
these magnetic properties by discussing results on electronic structures and magnetic exchange interactions.

\subsection{ Electronic structure of Mn$_{2}$NiX and dependencies on configurations}

Figs. 2-4 show a comprehensive comparative presentation of the densities of states in various
configurations for the three systems considered. In each figure, panels (a)-(d) show comparisons
of total and partial densities of states (of magnetic components) for {\bf OC} and {\bf DC} configurations.
Panels (e)-(h) of each figure show comparisons of same quantities for {\bf DC} and {\bf DT} configurations.
The comparisons between {\bf OC} and {\bf OT} configurations for these systems
had already been done \cite{paul}.

\begin{figure}[h]
\includegraphics[width=8.5cm]{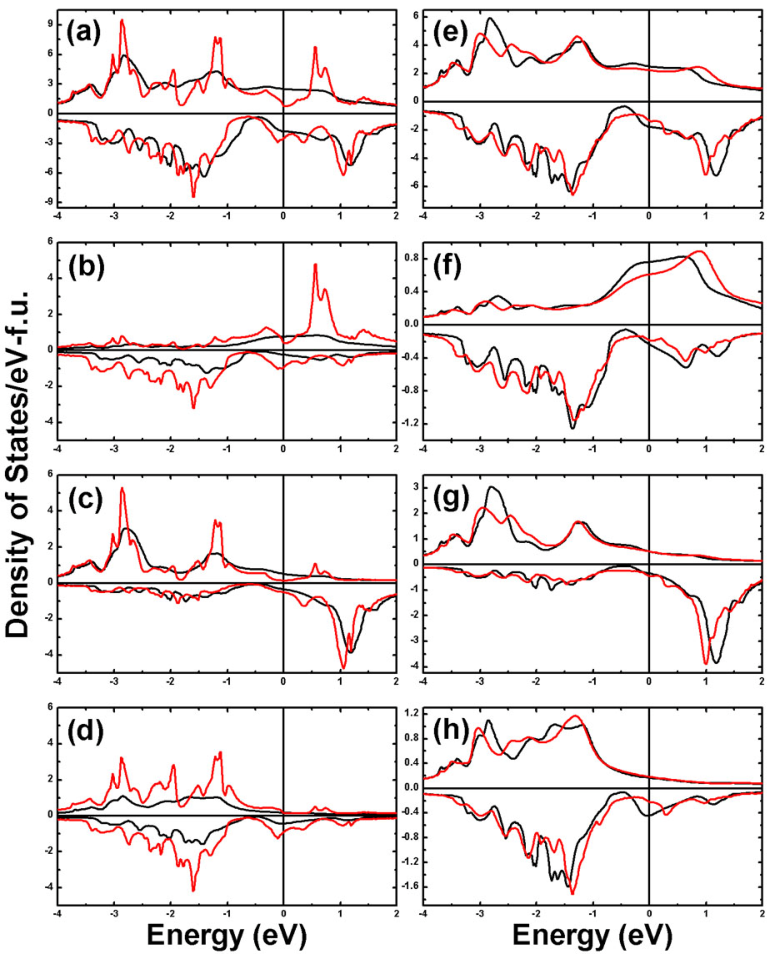}
\caption{(Color online) Total and partial densities of states for Mn$_{2}$NiGa. In panels (a)-(d) black curves stand for
{\bf DC} configuration and red curves stand for {\bf OC} configuration. In panels (e)-(h), black curves denote
the {\bf DC} configuration and red curves denote the {\bf DT} configuration. (a)-(d) and (e)-(h) panels display
total, MnI, MnII and Ni densities of states, respectively.}
\end{figure}

The densities of states of all three alloys in the {\bf OC} configuration show certain common features: the MnI densities
of states have characteristic unfilled majority bands as can be inferred by presence of substantial states in the
unoccupied part, while MnII have same characteristics associated with their minority bands. This explains the reasons
for getting substantial compensation of Mn moments leading to rather small total moments. The features in the minority
bands near the Fermi level emerge due to hybridizations between MnI and Ni 3$d$ states. For Mn$_{2}$NiGa and Mn$_{2}$NiIn,
such hybridizations give rise to a small peak around -0.25 eV in the minority bands while for Mn$_{2}$NiSn, a prominent
peak around -0.5 eV emerges.
The prominent peaks in the occupied part of the
minority bands such as the ones between -1 eV and -2 eV also arise from the hybridizations of same states. The
majority bands for all three systems too have common characteristics; the features in the occupied parts arise due to
hybridizations between Ni and MnII states.

\begin{figure}[h]
\includegraphics[width=8.5cm]{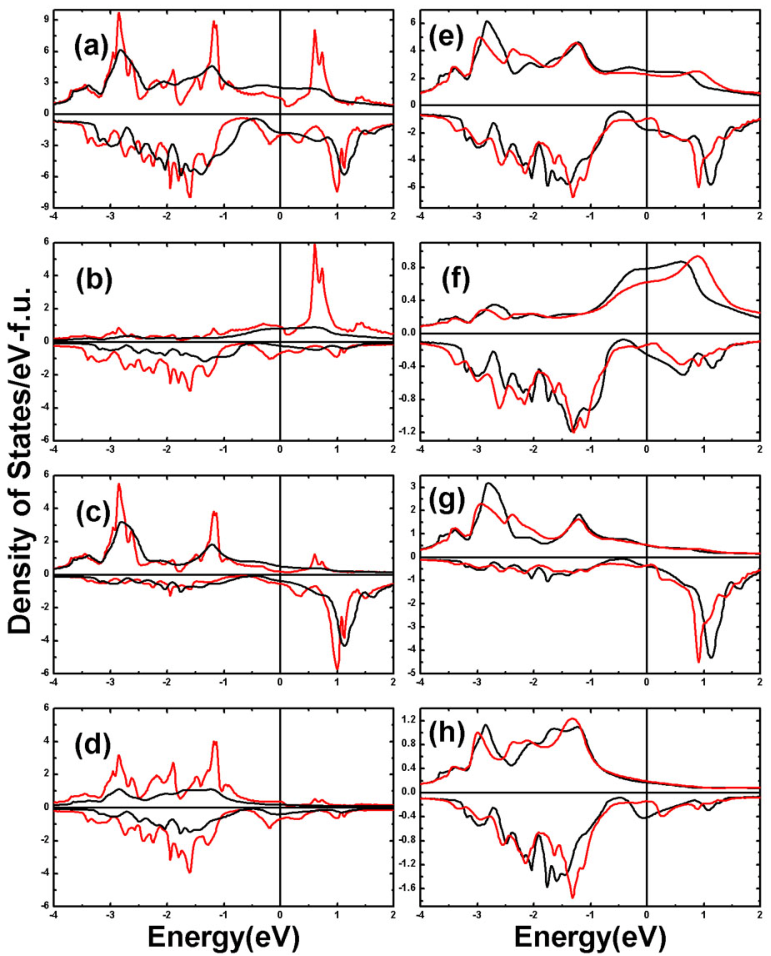}
\caption{(Color online) Total and partial densities of states for Mn$_{2}$NiIn. IIn panels (a)-(d) black curves stand for
{\bf DC} configuration and red curves stand for {\bf OC} configuration. In panels (e)-(h), black curves stand for
{\bf DC} configuration and red curves stand for {\bf DT} configuration. (a)-(d) and (e)-(h) panels display
total, MnI, MnII and Ni densities of states, respectively.}
\end{figure}

%\begin{table}[!htbp]
%        \caption{The inter-atomic distances (in $\AA$) between the magnetic
%        atoms for Mn$_{2}$Ni{\it X} systems in cubic and tetragonal phases.}
%        \begin{center}
%            \begin{tabularx}{.4\textwidth}{|c|c|c|c|c|c|c|} \hline
%                Systems & \multicolumn{3}{|c|}{Cubic} & \multicolumn{3}{|c|}{Tetragonal}\\ \hline
%                Mn$_{2}$Ni{\it X} &MnI-MnII &Ni-MnI &MnII-Ni &MnI-MnII &Ni-MnI &MnII-Ni \\ \hline
%                Mn$_{2}$NiGa      &2.56      &2.95   &2.56     &2.58      &2.77   &2.58 \\
%                Mn$_{2}$NiIn      &2.58      &2.98   &2.58     &2.61      &2.76   &2.61 \\
%                Mn$_{2}$NiSn      &2.64      &3.05   &2.64     &2.66      &2.87   &2.66 \\ \hline
%        \end{tabularx}
%    \end{center}
%\end{table}

Significant modifications to the densities of states in the {\bf OC} configurations occur due to anti-site
disorder between MnI and Ni sites. Expectedly, major changes in the electronic structure come from the MnI and
the Ni densities of states. The total densities of states in the majority spin channel become rather featureless
and flat in the {\bf DC} configurations. The densities of states in the minority bands, although still retain
some of the structures, but become smooth in general. For example, the peaks near the Fermi level in the minority
bands of Mn$_{2}$NiGa and Mn$_{2}$NiIn and the peaks at the Fermi level in the minority bands of
Mn$_{2}$NiSn are all broadened in the {\bf DC} configuration. MnI densities of states are affected most followed by
the Ni ones. The minority spin MnI peaks (around -1.5 eV for Mn$_{2}$NiGa and Mn$_{2}$NiIn,
and around -2 eV for Mn$_{2}$NiSn) broaden considerably and move closer to Fermi level (around -1.4 eV for Mn$_{2}$NiGa
and Mn$_{2}$NiIn, and around -1.3 eV for Mn$_{2}$NiSn). The peaks near or at the Fermi
level in the minority channels are now destroyed producing a continuously increasing densities of states near the Fermi
levels. The peaks in the unoccupied parts of the majority bands of MnI atoms are severely modified: their intensities
decrease considerably and they flatten out into broad plateaus extending into the occupied parts. Thus, the
occupied parts in the majority bands start to get filled with states. Consequently, the
magnetic moments of MnI atoms decrease in magnitudes as compared to the ones in the {\bf OC} configurations. Densities
of states in both spin channels associated with Ni too get modified with sharp peaks in {\bf OC} configurations getting
broadened in general. The states in the minority channels start to shift towards the Fermi level, while there is hardly
any shift in the states in the majority channels (comparison of panels (d) and (h) illustrate these clearly). These
features explain the slight increases in the Ni moment in the {\bf DC} configurations in comparison to the {\bf OC}
configurations of all three alloys. The anti-site disorder between MnI and Ni sites do not affect MnII densities of states
substantially.The major features of the MnII densities of states in {\bf OC} configuration do not change in {\bf DC}
configurations, the peaks only broaden explaining why the MnII moments remain almost intact in spite of anti-site disorder.
To summarize, the drastic re-distribution of electronic states in both spin channels of primarily MnI atoms, brought about
by the anti-site disorder between the sites of same point group symmetry, quenches the MnI moments substantially reducing
the exchange splitting and thereby increasing the total magnetic moment significantly as compared to the {\bf OC}
configurations.The degree of this increment depends on the degree of changes in the MnI densities of states. The maximum
changes occur in Mn$_{2}$NiSn, where, the peak
at -2 eV in the MnI minority band moves substantially towards the Fermi level, relocating itself at -1.25 eV as one
goes from {\bf OC} to {\bf DC}, the other peaks too make such a move, affecting the electron distribution considerably,
explaining why the changes in the moments are the largest in the series.

\begin{figure}[h]
\includegraphics[width=8.5cm]{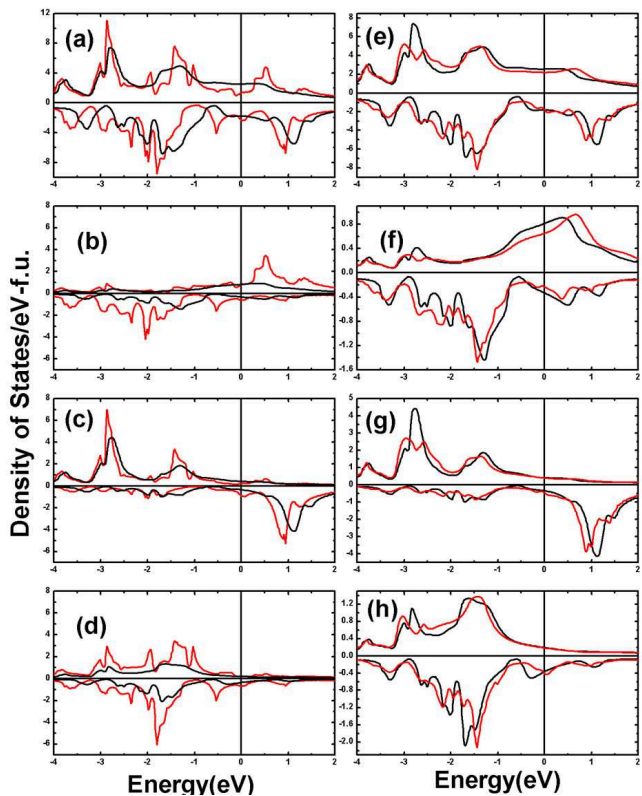}
\caption{(Color online) Total and partial densities of states for Mn$_{2}$NiSn. In panels (a)-(d) black curves stand for
{\bf DC} configuration and red curves stand for {\bf OC} configuration. In panels (e)-(h), black curves stand for
{\bf DC} configuration and red curves stand for {\bf DT} configuration. (a)-(d) and (e)-(h) panels display
total, MnI, MnII and Ni densities of states, respectively.}
\end{figure}

The reason behind obtaining a large $\Delta M$ when anti-site disorder is present, as compared to
the ``ordered'' configurations is that the magnetic moments do not
change as substantially as the systems undergo martensitic transformations. The percentage changes quoted
in Section IIIC exemplify this. In order to understand the reasons behind this, one needs to inspect the
panels (e)-(h) in FIG. 2-4 where comparisons between densities of states in the {\bf DC} and {\bf DT}
configurations are done. The results suggest that for all three materials, the total densities of states at
the Fermi level in the {\bf DT} configurations are less than that in the {\bf DC} configurations. This provides
a clue to the stability of the martensitic phases. In fact, the densities of states are suppressed prominently
in the minority bands of between -0.3 eV and the Fermi level in the {\bf DT}
configurations. In contrast, the minority densities of states
in {\bf DT}
configurations for all three materials get elevated as compared to the {\bf DC} configurations between -0.3 eV and -1 eV. These undoubtedly point to  the fact that electron states are transferred to the lower energies as the systems undergo martensitic transformations, explaining the stabilizations of the martensitic phases.

From the atom projected densities of states, it is clear that the hybridizations of Ni-MnI 3$d$ minority electrons are
responsible for such re-distributions of electronic states when the system undergoes a tetragonal deformation. Due to
the tetragonal distortion, the Ni-MnI hybridizations strengthen due to the significant reductions of the Ni-MnI bond
distances. Table V shows the bond distances between various Mn atoms in the austenite and in the martensitic phases.
The results clearly demonstrate that only the Ni-MnI inter-atomic distances reduce by almost $6 \%$ in all the cases.
On the other hand, the changes in the majority band electronic structures due to tetragonal distortions are overwhelmingly
due to the MnI states. A comparison between {\bf DC} and {\bf DT} configurations show that states in the MnI majority
bands are pushed into the unoccupied part as the systems undergo martensitic transformations. This coupled with the
shifting of minority states towards lower energies produce a larger exchange splitting of MnI resulting in an increase
of the net MnI moment as compared to the austenitic phases. Thus the total moment in the martensitic phases do
not change as much as they do in the austenite phases upon changes in the sublattice occupancies due to anti-site
disorder in the sites with tetrahedral symmetries, resulting in a larger $\Delta M$ in the ``disordered'' configurations.

\subsection{Exchange interactions and Curie temperatures of Mn$_{2}$NiX}

\begin{figure}[h]
\includegraphics[width=8.5cm]{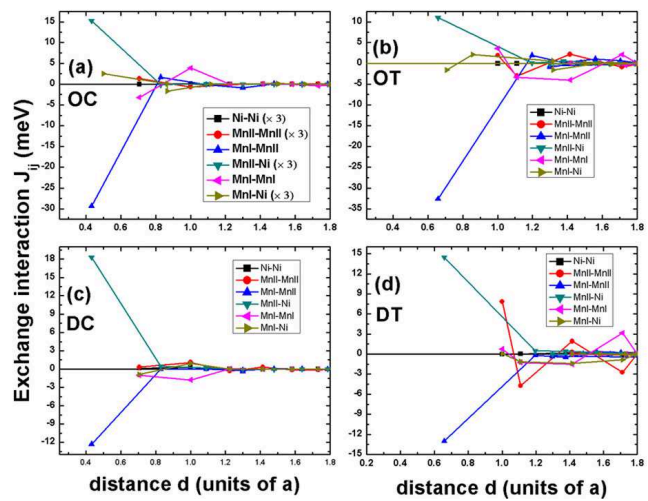}
\caption{(Color online) Magnetic exchange interactions (J$_{ij}$) as a function of inter-atomic distance d
for Mn$_{2}$NiGa. Panel (a), (b), (c) and (d) represent ordered cubic ({\bf OC}), ordered
tetragonal ({\bf OT}), disordered cubic ({\bf DC}) and disordered tetragonal ({\bf DT}) configurations,
respectively.}
\end{figure}

In this section, we show comparative results on inter-atomic exchange interactions across structures, configurations
and materials in order to understand the trends in the magnetic properties of Mn$_{2}$Ni{\it X} series. In FIG. 5-7 we
present results on inter-sublattice and intra-sublattice exchange interactions. The results suggest that in
all three materials, the magnetic properties are governed by competitions between two interactions: Ni-MnII and MnI-MnII.
The MnI-MnII interactions are anti-ferromagnetic while the Ni-MnII interactions are ferromagnetic. Comparisons of {\bf OC}
and {\bf OT} configurations show that upon tetragonal distortions, the anti-ferromagnetic MnI-MnII interactions strengthen
while the ferromagnetic Ni-MnII interactions weaken, resulting in a net loss of MnII moments and subsequent low moments
in the {\bf OT} configurations. Drastic modifications in the exchange interactions are observed due to anti-site
disorder. A comparison between {\bf OC} and {\bf DC} configurations show that for all three alloys, the Ni-MnII interactions
strengthen that is become more ferromagnetic while the MnI-MnII interactions weaken substantially. The weakening of the
most prominent anti-ferromagnetic interaction and strengthening of the most prominent ferromagnetic interaction in
presence of anti-site disorder, strengthen the ferromagnetic interactions in the systems resulting in the enhancement of
the overall magnetic moments when anti-site disorder is present. However, comparative assessments of {\bf DC} and {\bf DT}
configurations show that upon tetragonal distortions, the ferromagnetic Ni-MnII interactions weaken and the antiferomagnetic
MnI-MnII interactions strengthen.The weakening of the former being substantial, the anti-ferromagnetic interactions in
the {\bf DT} configurations are more significant than those in the {\bf DC} configurations, resulting in a lower total
moment in the former configurations in comparison to the later ones.

\begin{figure}[h]
\includegraphics[width=8.5cm]{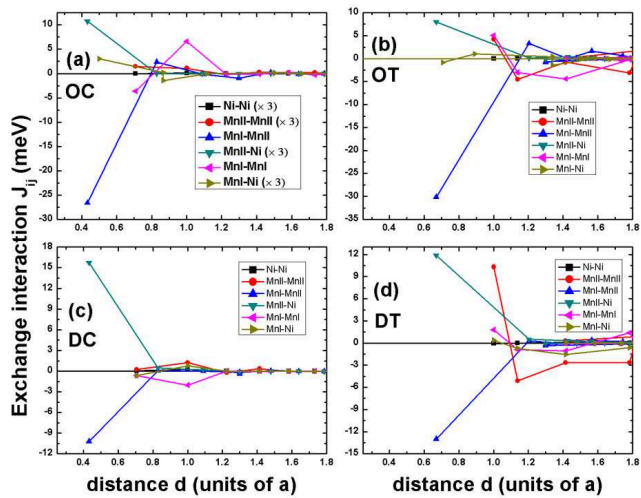}
\caption{(Color online) Magnetic exchange interactions (J$_{ij}$) as a function of inter-atomic distance d
for Mn$_{2}$NiIn. Panel (a), (b), (c) and (d) represent ordered cubic ({\bf OC}), ordered
tetragonal ({\bf OT}), disordered cubic ({\bf DC}) and disordered tetragonal ({\bf DT}) configurations,
respectively.}
\end{figure}

The intra-sublattice exchange interactions are much weaker than the inter-sublattice ones and thus do not contribute
enough to understand the trends in the magnetic properties. However, unlike the inter-sublattice ones, no common trend
across the materials is observed in some of these interactions. The Ni-Ni interactions are the weakest and are weakly
ferromagnetic for all materials and for all configurations. The MnI-MnI interactions are oscillatory in the {\bf OC}
configurations and become slightly anti-ferromagnetic in the {\bf DC} configurations. In the {\bf DT} configurations, they
again become oscillatory with first neighbor being strongly ferromagnetic and second and third neighbors being slightly
anti-ferromagnetic; the only exception being Mn$_{2}$NiSn where the second and third neighbor anti-ferromagnetic interactions
are stronger than the first neighbor ferromagnetic one. The Ni-MnI interactions become weak and predominantly anti-ferromagnetic
as one goes from {\bf OC} to {\bf DC} configurations where they are predominantly ferromagnetic. Under tetragonal
distortions in the {\bf DC} configurations
,the first neighbor interactions change to become slightly ferromagnetic offering
no other significant changes. The MnII-MnII interactions vary qualitatively quite a bit across materials. In the {\bf OC}
configurations, they are oscillatory in case of Mn$_{2}$NiSn with dominant interactions being
ferromagnetic. For the other two materials, the interactions are
primarily ferromagnetic and weaker in comparison to Mn$_{2}$NiSn. The anti-site disorder keeps the interactions largely
intact except that they are weaker in Mn$_{2}$NiSn. The tetragonal distortions modify these interactions significantly
by making them more oscillatory. However, the strengths of the ferromagnetic and anti-ferromagnetic components in the
oscillatory exchange interactions are nearly equal and thus compensate. To summarize, the magnetic exchange interactions
in these materials are thus influenced by the inter-sublattice interactions and changes in their relative strengths upon
changes in structures and configurations help make connections to the trends in the magnetic properties.

\begin{figure}[h]
\includegraphics[width=8.5cm]{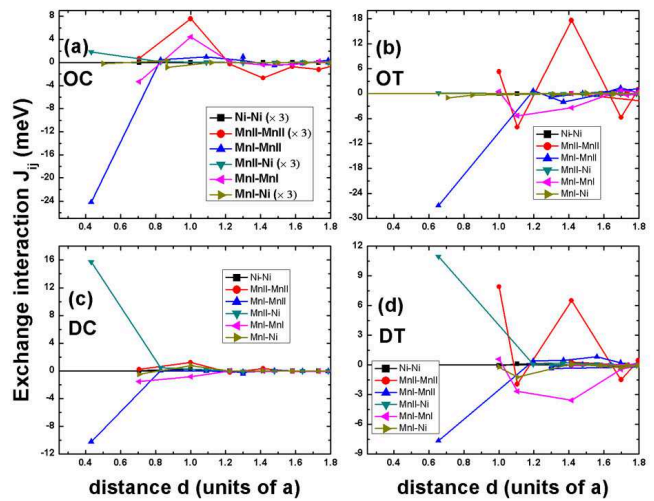}
\caption{(Color online) Magnetic exchange interactions (J$_{ij}$) as a function of inter-atomic distance d
for Mn$_{2}$NiSn. Panel (a), (b), (c) and (d) represent ordered cubic ({\bf OC}), ordered
tetragonal ({\bf OT}), disordered cubic ({\bf DC}) and disordered tetragonal ({\bf DT}) configurations,
respectively.}
\end{figure}

\section{Conclusions}

By means of {\it ab initio} calculations, we have investigated compositional and structural stabilities along with the magnetic
properties of inverse Heusler Mn$_{2}$Ni{\it X} alloys in connection to the improvement of  shape memory and magneto-caloric effects.
We find that the three out of four alloys considered here are potentially better shape memory materials  as a large
value of $\Delta M$ can be achieved even in these compositions whereas in prototype
Ni$_{2}$Mn{\it X} systems, one has to vary the ratio of Ni, Mn and {\it X} judiciously to achieve the same. We also find that these
materials would show inverse magneto-caloric effect, which is technologically desirable for green environment. Our
analysis reveal that these properties emerge due to the inverse Heusler structure where the two Mn atoms are
inequivalent and an anti-site disorder exists between one of the Mn and the Ni sublattices. It is also found that the
electronic structures associated with the Mn atom, which makes an alloy with Ni, are primarily responsible for the dramatic
changes in the magnetic properties, and consequently for improved functionalities. This work, thus, show that
the Mn$_{2}$Ni{\it X} alloys in the inverse Heusler
structure can be considered as potential functional materials and that more experimental verifications of their
functional properties are required.

\section{Acknowledgments}

Financial assistance from the Swedish Research Links (VR-SIDA) is acknowledged. The
Swedish National Computing facilities, computation facilities from C-DAC, Pune, India
and from Department of Physics, IIT Guwahati funded under the FIST programme of DST, India
are also acknowledged.

\end{document}